# DESIGNING PROGRAMMING EXERCISES FROM BOARD GAMES

Maxim Mozgovoy and Marina Purgina
School of Computer Science and Engineering
The University of Aizu
Tsuruga, Ikki-machi, Aizuwakamatsu, Fukushima
Japan
E-mail: mozgovoy@u-aizu.ac.jp, mapurgina@gmail.com

## ABSTRACT

This paper introduces a collection of board games specifically chosen to serve as a basis for programming exercises. We examine the attractiveness of board games in this context as well as features that make a particular game a good exercise. The collection is annotated across several dimensions to assist choosing a game suitable for the target topic and student level. We discuss possible changes into exercise tasks to make them more challenging and introduce new topics. The work relies on established topics taxonomy and board games resources which makes extending the current collection easy.

## INTRODUCTION

Games commonly serve as a good starting point for programming exercises [1, 2]. Such exercises offer a unique combination of attractive features ranging from diverse technical skills required to complete them to user interface and artistic challenges on the way to appealing results. Games are often used as example environments in a variety of domains, including computer graphics, sound, AI, and networking. Games are included into collections of sample projects like Rosetta Code or distributed with development tools (Microsoft QuickBasic, Visual Studio).

While the complexity and diversity of games as software systems is evident, and their high potential as exercise projects is obvious, introducing games into a regular entry-level programming course is not always easy. Even a simple game might demand considerable programming skills and knowledge of specialized topics such as computer graphics and GUI design. It is also easy to misjudge the complexity of a project and its scope just by reading game description. A game project might require the student to spend disproportionate amount of time on marginal tasks like data entry or face advanced topics like making AI.

Therefore, potential game project exercises have to be carefully evaluated in advance to understand their real complexity and suitability for a particular group of students. The goal of this paper is to introduce a small but growing collection of board games, handpicked specifically as exercises for beginner software developers[1]. We will discuss the rationale behind game selection process, typical programming challenges arising in games, and suggest possible extensions of the proposed projects.

## RELATED WORK

The present paper owes much to the work by Drake and Sung [3], where an earlier attempt to collect games suitable as programming exercises is introduced. The authors discuss a variety of topics arising in this activity, and their reflections on possible criteria for inclusion of particular games are especially valuable. Here we will address many of the same questions as raised by Drake and Sung.

Numerous authors describe their experience of introducing a *particular* game into course curriculum. Choices vary from classics such as Chinese checkers and Mancala [4] to modern strategies like Ticket to Ride [5] and custom-made games, designed for a specific purpose, e.g., teaching AI [6].

As already noted, games offer a diverse variety of challenges to a programmer, so it is easy to imagine the use of a particular game at any specialized course. It is harder to propose a systematic approach for choosing games, suitable for specific teacher and learner needs. This problem is closely related to the task of classifying programming exercises in general: an exercise should be accompanied with certain information helping the student and/or teacher understand its objective, scope, and complexity.

There are attempts to develop such classification schemes, emphasizing different aspects of the educational process. For example, Fuller et al. [7] focus on student abilities, evaluating the relevance of a particular exercise on the basis of student's skills of *interpreting* and *producing* computer programs. Such type of taxonomy is useful for adapting the given exercise to a certain learner level, but it speaks less about intrinsic properties of the software system we are aiming to produce. A more relevant approach type for our purposes is proposed by Santos et al. [8], who classify each exercise across three dimensions:

1. *Topics*. The list of common topics in a typical introductory programming course, such as "variables and operators", "conditional structures", "functions", etc. An exercise can be classified as belonging to several topics.
2. *Complexity*. A certain "complexity score" of an exercise. If necessary, several different complexity types can be assigned, such as "math complexity", "code complexity", or "cognitive effort".

---

[1] https://github.com/rg-software/board-games

3. *Levels*. A list of levels assigned within alternative models. The authors suggest providing a Bloom's taxonomy level and an intended student type (beginner, intermediate, advanced).

Note that the dimensions above specify the general classification framework rather than precisely defined categories to be used. Thus, it has to be adapted for our needs.

**SHORTLISTING GAMES**

Following the example of Drake and Sung [3], the present work focuses specifically on *board games*, defined by these authors as *"board, card, or dice games that are typically not played on a computer"*. The largest online resource and community of board game enthusiasts BoardGameGeek[2] (BGG) gives no definition of a board game, instead providing a lengthy list of games and game-like activities considered *outside* the scope of the site[3]. Typical examples of board games include chess, poker, snakes and ladders, etc.

Most board games share certain features, making them attractive as exercises, especially in comparison with typical video games:

1. Board games are generally turn based, which makes complex subsystems of animation, physics, and real-time player control unnecessary or optional.
2. A complete rulebook of any board game is known in advance, so there is no need to "reverse engineer" game logic, as may happen even with simple video games.
3. Board games do not require fast reflexes or other dexterity skills from the player, which makes them accessible to a wider audience.

Almost any board game project can be easily adjusted for the desired complexity by means of gradual inclusion of subsequent optional elements. For example:

1. (Base project). Implement a text-based version of the system, allowing the minimal required number of human players to complete one game session.
2. Implement a system of unit tests for the game (can be a part of the base project if using test-driven development).
3. Implement a GUI for the game.
4. Implement support for additional players, if applicable.
5. Make the game resettable, i.e., allow to play another session after the first one is finished.
6. Make the game saveable: let the users save the current session to a file and reload it later.
7. Implement a remote (network) play capability.
8. Implement a game AI system.

Even the base project can be made simpler, for example, by making illegal moves protection less strict or optional. It is also possible to provide ready unit tests and ask the students to implement code to pass them.

Since our goal is to use games as entry-level exercise projects, each game has to pass a certain "filter" evaluating its viability in this context. Following the guideline of [3], we can propose its extended and modified version:

1. The game should be relatively quick to play, ideally under 15 minutes. Game rules should be short, clear, and easy to implement in code.
2. The game should be designed for 2+ players (which makes it a good pair programming exercise), but occasional deviations are acceptable.
3. The game should presume that the same information is available to all the players, which makes playing on the same shared computer possible. This restriction excludes most card games, however.
4. The game should be considered within the scope of BGG and have a dedicated BGG page. This requirement makes it easy to find more information about the game, including its possible expansions and variations.
5. The previous condition excludes *solo puzzles*, such as 15 puzzle or Rush Hour. They are considered different from solo or *de facto* solo games like card solitaire or Mastermind, since they generally come with a list of predefined problems and do not introduce randomicity that makes each game session unique.
6. Game coding should not involve long tedious tasks (such as typing lists of card effects or drawing custom boards).
7. The game should not include language-dependent elements, making them harder to reuse in an international setting (which excludes most word games).
8. The game should be at least *mildly* engaging, which can be defined as "having an average BGG rating of 5 out of 10 or higher" for our purposes.

Some of these features are correlated. For example, quick games tend to have simpler rules. However, simple games are typically not found among the most high-ranked BGG entries.

The last decades were marked with the arrival of new generations of innovative board games, making older games look less attractive: most games on BGG with the user rating of 8 and higher and at least 50 reviews are released after 2000. Drake and Sung note that newer games are more likely to be covered by some kind of legal protection (copyright and trademarks, and far less commonly patents).

Relevant sources, such as [9–11] suggest that at least in the North American context legal protection typically covers trademarked names, visual art, and the exact wording of a rulebook. Thus, one should be careful about reusing the original text and graphics of a game. However, implementing core game mechanics as a programming exercise is not likely to cause legal issues according to the referenced sources and to the best of authors' knowledge. Note that the authors are not lawyers and cannot provide legal advice.

**DEVELOPING COLLECTION ATTRIBUTES**

Once the criteria for choosing games are set, we need to review the taxonomy of Santos et al. [8] and adapt it for our purposes.

---

[2] https://boardgamegeek.com

[3] https://boardgamegeek.com/boardgame/23953/outside-scope-bgg

The dimension of *Topics* is perhaps the most challenging as the notion of "common topics in an introductory programming course" is vague. An attempt to provide a short list of topics and evaluate their difficulty to the student is made by Meisalo et al. [12]. They identify the following items: Variables and symbols, Input and output, Conditional statements, Loops, Arrays, Methods, [Java] Applets, Graphics, Key Event, Animations. Other lists can be compiled by examining the contents of good introductory books on programming. For example, a classic "K&R" book [13] contains the chapters titled "Types, operators, and expressions", "Control flow", "Functions and program structure", "Pointers and arrays", "Structures", and "Input and output". It may be argued that the choice of chapters depends on a particular language used. For example, a Python-based book [14] has a chapter on "Loops and lists", placing them into the same category. Certain topics like "classes" or "recursion" are hard to assign since any project can be completed without these instruments.

Table 1: *Topic* Dimension Categories

| Topic | Comment |
| --- | --- |
| Basics | Assignments, simple branches and loops. |
| Arrays | One-dimensional arrays and lists. |
| 2D Arrays | Two-dimensional arrays |
| Algorithms | Basic algorithms (searching, sorting, etc.) |
| Algorithms+ | More advanced algorithms like matrix transposition and/or tricky techniques. |
| Graphs | Graph representations and algorithms |

The dimension of *Complexity* can be roughly estimated by the required lines of code (LOC) used to implement the core functionality of the game (without user interface). While some projects are short but tricky to implement and vice versa, all of them ultimately belong to the "simple board game" type, so

Table 2: The present content of the board games collection (sorted according to LOC)

| Game | BGG ID[*] | BGG Rating | Core LOC | GUI Value | Players | Category | Topics |
| --- | --- | --- | --- | --- | --- | --- | --- |
| Pig | 161130 | 5.3 | 25 | Low | 2 | Dice | Basics |
| Mastermind | 2392 | 5.6 | 25 | Low | 1-2 | Deduction | Basics, Arrays |
| GOLO (basic) | 7270 | 5.6 | 25 | Low | 1+ | Dice | Basics, Arrays |
| Kalah | 2448 | 5.9 | 50 | Low | 2 | Abstract | Arrays |
| Stop-Gate | 7450 | 6.1 | 50 | High | 2 | Abstract | 2D Arrays |
| No Thanks! | 12942 | 7.1 | 50 | Low | 3-7 | Cards | Arrays, Algorithms |
| Othello | 2389 | 6.1 | 50 | High | 2 | Abstract | 2D Arrays, Algorithms+ |
| Impact | 246228 | 6.7 | 50 | Low | 2-5 | Dice | Arrays, Algorithms |
| Gold Fever | 234120 | 6.4 | 50 | Low | 2-5 | Cards | Basics, Arrays |
| GOLO (scorecard) | 7270 | 5.6 | 50 | Low | 1+ | Dice | Arrays, Algorithms |
| Ship, Captain, and Crew | 18812 | 5.1 | 50 | Low | 2+ | Dice | Arrays, Algorithms |
| Quixo | 3190 | 6.2 | 50 | High | 2-3 | Abstract | Arrays, Algorithms |
| Poker dice | 10502 | 5.1 | 100 | Low | 2+ | Dice | Arrays, Algorithms |
| Paletto | 101463 | 6.7 | 100 | High | 2-3 | Abstract | Graphs, Algorithms+ |
| Black Box | 165 | 6.4 | 100 | Low | 1-2 | Deduction | 2D Arrays, Algorithms+ |
| Criss Cross | 220988 | 6.4 | 100 | High | 1-6 | Dice | 2D Arrays, Algorithms |
| King's Valley | 86169 | 6.5 | 100 | High | 2 | Abstract | 2D Arrays, Algorithms |
| Farmers Finances | 201028 | 6.3 | 150 | Low | 2 | Economic | Basics |
| Orchard | 245487 | 7.4 | 150 | High | 1 | Cards | 2D Arrays, Algorithms+ |
| Blokus Duo | 16395 | 6.8 | 200 | High | 2 | Abstract | 2D Arrays, Algorithms+ |
| Push Fight | 54221 | 7.4 | 200 | High | 2 | Abstract | 2D Arrays, Graphs |

[*]Check the BGG game page at https://boardgamegeek.com/boardgame/<BGG ID>

Clearly, some of these topics like "input and output" are relevant for *all* board games. Others like "animations" or "graphics" are deliberately made optional during initial task definition. Thus, both these topic types are irrelevant for us. It is also evident that some topics appear especially often in board games. A typical *board* can be represented as a rectangular matrix, so "two-dimensional arrays" will probably be the most common topic. The currently used selection of topics is provided in Table 1.

a simple line count is probably more reliable than any subjective score assigned by the author.

The *Levels* dimension as described in Santos et al. [8] is hardly applicable to our case. The target Bloom level is always 6 ("creating") [15], and the target student level is roughly the same for all the projects.

There are, however, other attributes that can be useful for our context. They are *Game category*, *Number of players*, and

*GUI value*. The last attribute (low/high) indicates whether the game is playable without a graphical interface, or making GUI is strongly desirable. These attributes along with the game's BGG rating constitute the complete markup of each game in the collection (see Table 2).

## IMPLEMENTATION DETAILS

Each game in the collection is implemented according to the "stage 3" complexity of the task: there are both text (dialog-based) and graphical user interface versions, and a reasonable set of unit tests. Implementations are supposed to be simple and straightforward, resembling typical solutions. All games are coded in Python. Being concise, Python provides a good basis for the lower-bound LOC estimation. Graphical user interface is created using Pygame Zero[4], which adds minimal overhead to text-based implementations. No other third-party libraries are used.

It may be argued that providing reference implementations for all the games in the collection tempts the students to "borrow" existing code. While it might be true, it also greatly reduces the burden of a teacher willing to design a simpler "fill the gaps" exercise by providing an option to remove some parts of the code. It also makes easy to check student solutions against the most probable source of their "borrowings"

## DISCUSSION AND CONCLUSION

The main goal of this work is to create a small-scale annotated catalogue of board games, suitable as programming assignments, and provide sample implementations. Using this catalogue, it should be possible to identify games of desired scale and complexity and adapt them to specific needs of a particular course. The author, for instance, successfully uses some of these games at a Concurrent and Distributed Systems course, where typical assignments presume the implementation of remote play functionality.

The stated filtering criteria allowed to choose games equally applicable in a variety of contexts, so the teachers and learners can focus on the target topic and complexity rather than on individual peculiarities of a particular game. Unfortunately, the proposed filtering favors certain game genres, especially abstract strategy. Hopefully, the inclusion of new games will increase genre diversity.

## REFERENCES

[1] D. C. Cliburn, "The effectiveness of games as assignments in an introductory programming course," in *Proceedings. Frontiers in Education. 36th Annual Conference*, 2006, pp. 6–10.

[2] K. Sung, "Computer games and traditional CS courses," *Commun. ACM*, vol. 52, no. 12, pp. 74–78, 2009, doi: 10.1145/1610252.1610273.

[3] P. Drake and K. Sung, "Teaching introductory programming with popular board games," in *SIGCSE'11: Proceedings of the 42nd ACM Technical Symposium on Computer Science Education, March 9-12, 2011, Dallas, Texas, USA*, New York, New York, USA, 2011.

[4] T. Huang, "Strategy game programming projects," *Journal of Computing Sciences in Colleges*, vol. 16, no. 4, pp. 205–213, 2001.

[5] D. Lim, "Taking students out for a ride: using a board game to teach graph theory," *ACM SIGCSE Bulletin*, vol. 39, no. 1, pp. 367–371, 2007.

[6] D. Ashlock, J. A. Brown, C. Gregor, and M. Makhmutov, "A Family of Turn Based Strategy Games with Moose," in *2021 IEEE Symposium Series on Computational Intelligence (SSCI)*, Orlando, FL, USA, 2021, pp. 1–8.

[7] U. Fuller *et al.*, "Developing a computer science-specific learning taxonomy," *ACM SIGCSE Bulletin*, vol. 39, no. 4, pp. 152–170, 2007.

[8] Á. Santos, A. Gomes, and A. Mendes, "A taxonomy of exercises to support individual learning paths in initial programming learning," in *2013 IEEE Frontiers in Education Conference (FIE)*, 2013, pp. 87–93.

[9] E. Sargeantson, *How to Protect Board Games with Copyrights, Patents and Trademarks.* [Online]. Available: https://mykindofmeeple.com/protect-board-game-copyrights-patents-trademarks/

[10] J. Bailey, "The Rise of Board Game Plagiarism," *Plagiarism Today*, 24 Jul., 2018. https://www.plagiarismtoday.com/2018/07/24/the-rise-of-board-game-plagiarism/

[11] D. J. Schaeffer, "Not Playing Around: Board Games and Intellectual Property Law," *Landslide*, vol. 7, no. 4, 2015. [Online]. Available: https://www.americanbar.org/groups/intellectual_property_law/publications/landslide/2014-15/march-april/not-playing-around-board-games-intellectual-property-law/

[12] V. Meisalo, E. Sutinen, and S. Torvinen, "Classification of exercises in a virtual programming course," in *34th Annual Frontiers in Education, 2004. FIE 2004*, 2004, S3D-1.

[13] B. W. Kernighan and D. M. Ritchie, *The C programming language,* 2nd ed. Englewood Cliffs, N.J.: Prentice Hall, 1988.

[14] H. P. Langtangen, *A primer on scientific programming with Python,* 5th ed. Berlin: Springer, 2016.

[15] D. R. Krathwohl, "A revision of Bloom's taxonomy: An overview," *Theory into practice*, vol. 41, no. 4, pp. 212–218, 2002.

---

[4] https://pygame-zero.readthedocs.io